\documentclass[aps,prb,preprint,showpacs,preprintnumbers,amsmath,amssymb,floatfix]{revtex4}

\usepackage{amssymb,amsmath}
\usepackage{graphicx}
\usepackage{bm}
\usepackage{dcolumn,epsfig}
\usepackage{color}

\newcommand{\beq}{\begin{equation}}
\newcommand{\eeq}{\end{equation}}
\newcommand{\bes}{\begin{subequations}}
\newcommand{\ees}{\end{subequations}}
\newcommand{\bea}{\begin{eqnarray}}
\newcommand{\eea}{\end{eqnarray}}
\newcommand{\ba}{\begin{array}}
\newcommand{\ea}{\end{array}}
\newcommand{\beqn}{\begin{eqnarray*}}
\newcommand{\eeqn}{\end{eqnarray*}}

\newcommand{\f}[2]{\frac{#1}{#2}}

\def\nn{\nonumber}

\newlength{\sizeonefig}
\newlength{\sizetwofig}
\begin{document}

\title{The role of lighter and heavier embedded nanoparticles on the thermal conductivity of SiGe alloys}

\author{A. Kundu$^{1}$, N. Mingo$^{1,2}$, D. A. Broido$^3$, and D. A. Stewart$^4$} 
\address{$^1$ CEA-Grenoble, 17 Rue des Martyrs, Grenoble 38000, France}
\address{$^2$ Department of Electrical Engineering, University of California, Santa Cruz, California 95064, USA}
\address{$^3$ Department of Physics, Boston College, Chestnut Hill, Massachusetts 02467, USA}
\address{$^4$ Cornell Nanoscale Facility, Cornell University, Ithaca, New York 14853, USA}

\begin{abstract}
We have used an atomistic {\it ab initio} approach with no adjustable parameters to compute the lattice thermal conductivity of Si$_{0.5}$Ge$_{0.5}$ with a low concentration of embedded Si or Ge nanoparticles of diameters up to 4.4 nm. Through exact Green's function calculation of the nanoparticle scattering rates, we find that embedding Ge nanoparticles in $\text{Si}_{0.5}\text{Ge}_{0.5}$ provides 20\% lower thermal conductivities than embedding Si nanoparticles. This contrasts with the Born approximation which predicts an equal amount of reduction for the two cases, irrespective of the sign of the mass difference. Despite these differences, we find that the Born approximation still performs remarkably well, and it permits investigation of larger nanoparticle sizes, up to 60 nm in diameter, not feasible with the exact approach.
\end{abstract}

\vspace{0.5cm}

\maketitle
\section{Introduction}
Nanoparticle Embedded in Alloy Thermoelectric (NEAT) materials have been proposed as a means of improving the thermoelectric properties of solid solutions. In a clear experimental demonstration of this concept, Kim et al. found a remarkable reduction of the thermal conductivity ($\kappa$) of InGaAs upon the introduction of lattice matched ErAs nanoparticles, below the alloy limit, without any decrease of the thermoelectric power factor \cite{KimShakouri}. In some cases nanoparticles may also play an active role in increasing the power factor \cite{Zebarjadi}. Naturally forming nanoinclusions are also thought to be at the core of the rather high thermoelectric figure of merit (ZT) of LAST (Lead-Antimony-Silver-Tellurium) materials \cite{Kanatzidis}. Theoretically, the introduction of nanophases inside alloys to reduce $\kappa$ and improve ZT has been investigated using various approaches \cite{Klemens,SlackHussain,KimMajumdar,MingoNanolett}. However, these approaches approximate wave scattering via a continuum description, and they rely on adjustable parameters. Our aim is to overcome these drawbacks by performing a parameter free atomistic calculation, which includes the nanoparticle scattering to all orders. We will show that the latter has a large influence on the results, which become highly asymmetric with respect to the sign of the scatterer's mass difference. (First order perturbation theory yields a quadratic, symmetric dependence.) \textit{Thus the question of whether it is better to embed lighter or heavier nanoparticles becomes very relevant in the light of the full order calculation.} We provide the answer in the specific case of SiGe alloys, and we discuss its implications when developing novel NEAT materials.

Our parameter free approach to compute thermal conductivity is based on an exact numerical solution of the linearized Boltzmann Transport Equation (BTE) for phonons \cite{Sparavigna,Broido07,Ward09,Lindsay}.  As a result of the computational challenge, many works still circumvent this solution by resorting to a host of approximations, most notably the relaxation time approximation \cite{Ziman}. A few years ago some of us showed that it is possible to predict the lattice thermal conductivity of group IV single crystal semiconductors from first principles, i.e. using the fundamental physical constants as the sole experimental inputs \cite{Broido07,Ward09}.
In the present paper we extend our approach to the case of disordered solid solutions like Si$_x$Ge$_{1-x}$, including also embedded nanoparticles. This requires the {\it ab initio} calculation of elastic scattering rates due to both alloy disorder, and due to the nanoparticles. Here we use atomistic Green's function techniques to compute those rates to all orders, beyond the Born approximation. We will show that the full result may deviate noticeably from the Born approximation.

\section{Theory}
The thermal conductivity of a bulk material can be calculated as
\bea
\kappa = {1\over k_B T^2}{V_{\text{uc}}\over 8 \pi^3}\sum_{\lambda}{n_0(n_0+1)|v^z_\lambda|^2\hbar^2\omega_\lambda^2\tau_\lambda},
\eea
where $V_{uc}$ is the unit cell volume, the summation sign is a shorthand for $\sum_{\lambda'} \equiv \sum_{\alpha'}\int_{BZ}{d{\vec q'}}$ (the integral is performed over the volume of the Brillouin zone), and $\lambda$ stands for the phonon branch index and wavevector, $\{\alpha,\vec q\}$. $\omega_\lambda$ and $v^z_\lambda$ are, respectively, the frequency and the group velocity along the z-direction of the corresponding phonon and $n_0$ is their occupation number. The $\tau_{\lambda}$ are scattering times that contain all the information about the non equilibrium phonon distribution.
Details on the BTE and its solution have been given in \cite{Sparavigna, Broido07, Ward09, Lindsay}. We will just summarize it briefly. The equation to solve is
\begin{eqnarray}
\tau_\lambda = \tau_\lambda^0 + \tau_\lambda^0\Delta_\lambda,
\label{taulambda}
\end{eqnarray}
where 
\bea
\Delta_\lambda\equiv \sum_{\lambda'\lambda''}^+{\Gamma^+_{\lambda\lambda'\lambda''}(\xi_{\lambda\lambda''}\tau_{\lambda''}-\xi_{\lambda\lambda'}\tau_{\lambda'})}+\nonumber\\
\sum_{\lambda'\lambda''}^-{{1\over 2}\Gamma^-_{\lambda\lambda'\lambda''}(\xi_{\lambda\lambda''}\tau_{\lambda''}+\xi_{\lambda\lambda'}\tau_{\lambda'})}
+\sum_{\lambda'}{\Gamma_{\lambda\lambda'}\xi_{\lambda\lambda'}\tau_{\lambda'}},\label{deltalambda}
\eea
where, the $\Delta_{\lambda}$ term takes into account coupling of non-equilibrium $\lambda$ phonon modes to other phonon modes ($\lambda',\lambda''$) based on energy and momentum conservation, and 
\bea
1/\tau^0_\lambda\equiv \sum_{\lambda'\lambda''}^+{\Gamma^+_{\lambda\lambda'\lambda''}}+\sum_{\lambda'\lambda''}^-{{1\over 2}\Gamma^-_{\lambda\lambda'\lambda''}}
+\sum_{\lambda'}{\Gamma_{\lambda\lambda'}}.
\label{Eq:tau0}
\eea
The (+) and (-) symbols over the sums in Eq.~\ref{Eq:tau0} indicate sums over $\lambda',\lambda''$ for the two types of three phonon processes available, $\lambda  \pm \lambda' \leftrightarrow \lambda''$.
The meanings of the three-phonon terms, $\Gamma^+_{\lambda\lambda'\lambda''}$, and $\xi_{\lambda\lambda''}$  are given in Ref.~\onlinecite{Lindsay}.
The $\tau_{\lambda}$ were obtained through iterative solution of the BTE. For simplicity, the last term in $\Delta_\lambda$ (Eq.~\ref{deltalambda}) has been neglected. This term vanishes for the nearly isotropic elastic scattering of low frequency phonons, which dominate the thermal conductivity.
The effect of elastic scattering from nanoparticles and alloy disorder is included in the sum of $\Gamma_{\lambda\lambda'}$ in Eq.~\ref{Eq:tau0}.
 
The main effect of a Si or Ge impurity or nanoparticle on lattice vibrations is through its mass difference compared to that of the host lattice. 
The dynamical equation for the displacements $u_i$ is $\omega^2 {\bf M}{\bf u}={\cal K}{\bf u}$, where ${\cal K}_{ij}={\partial^2E\over\partial u_i\partial u_j}$ is the interatomic force constants matrix, and $M_{ij}=M_i\delta_{ij}$ is the mass diagonal matrix. When substitutional impurities of different mass are inserted, a diagonal matrix perturbation is added to the equation as ${\cal V}=-\big({\bf M'-M}\big)\omega^2$ where ${\bf M'}$ represents mass matrix of the impure system. This perturbation is non zero only on the degrees of freedom associated with the scatterer. In the practical solution of the problem we work with the mass normalized matrices, ${\bf K}\equiv {\bf M}^{-1}{\cal K}$, ${\bf V}\equiv -{\bf M}^{-1}{\cal V}$. It has been shown that the other perturbation term, corresponding to the differences in force constants, ${\cal K'-K}$, has a much lesser effect on the thermal conductivity of SiGe alloys (about 10\% of the total\cite{Abeles}.) In the case of nanoparticles, an estimation of its order of magnitude can be easily made. Rayleigh scattering due to mass difference or to differences in the elastic constants have similar expressions, except for the prefactor, $S^2$. In the first case, this prefactor goes as $S_{\mbox{\small{mass}}}\sim(M_n-M_m)/M_m$. In the second case it is $S_{\mbox{\small{el}}}\sim\left({v_{m}^2-{M_{n}\over M_{m}}v^2_n\over v^2_m}\right)$, where sub indexes $m$ and $n$ stand for matrix and nanoparticle respectively, and $v$ is the speed of sound. The presence of the ${M_{n}\over M_{m}}$ term is needed to ensure that $S_{\mbox{\small{el}}}$ is zero if only the atomic masses change, but not the IFC's. For Si$_{0.5}$Ge$_{0.5}$ we have $S^2_{\mbox{\small{mass}}}=0.4$, whereas $S^2_{L,\mbox{\small{el}}}\sim 0.01$ for L acoustic modes, and $S^2_{T,\mbox{\small{el}}}\sim 0.0004$ for transverse ones. Thus, the effect of different elastic constants is much smaller than that due to mass differences. Strain effects on the nanoparticle due to lattice mismatch could also induce additional scattering. In such case, a factor of order $\sim\gamma (a_n-a_m)/a_m$ needs to be added to  $S_{\mbox{\small{el}}}$, where $\gamma\sim 1$ is the Gr\"uneisen constant, and $a$ are the lattice constants of the nanoparticle and matrix materials respectively\cite{Klemens55}. Addition of this term does not change the order of magnitude of $S^2_{\mbox{\small{el}}}$, which remains smaller than $\sim 0.02$ and can be considered negligible compared with the mass difference effect. Taking into account the strain and IFC difference effects in an ab initio calculation would require the self consistent atomic relaxation of extremely large supercells, several times the size of the nanoparticles considered here. Therefore, given these difficulties, and the minor resulting effect on the total thermal conductivity, the IFC differences will be neglected here.

The exact elastic scattering amplitudes due to a random distribution of independent scatterers in a homogeneous medium is $\Gamma_{\lambda\lambda'}\equiv \sum_p f^p\Gamma^p_{\lambda\lambda'}$ with
\begin{eqnarray}
\Gamma^p_{\lambda\lambda'}={\Omega\pi\over2\omega^2}{1\over V_{p}}\big|\langle \lambda|\bf{T_p}(\omega^2)|\lambda'\rangle\big|^2\delta(\omega-\omega'),
\end{eqnarray}
where $f^p$ is the volume fraction of scatterers of type $p$, $V_p$ is the scatterer's volume, $\Omega$ is the volume into which the phonon eigenstates $|\lambda\rangle$ are normalized, and $\bf{T_p}(\omega^2)$ is the T matrix associated with the scatterer of type $p$\cite{Mingo10}. We have adopted a virtual crystal approximation (VCA) model for the medium, where the interatomic force constants and atomic masses of pure Si and Ge crystals are averaged according to their relative concentrations in the alloy. The total alloy scattering for bulk Si$_x$Ge$_{1-x}$ is given by the concentration weighted sum of the scattering probabilities of a Si impurity in the VCA medium, and a Ge impurity in the VCA medium: 
$\Gamma^{\text{SiGe}}_{\lambda\lambda'}=x\tilde{\Gamma}^{\text{Si}}_{\lambda\lambda'}+(1-x)\tilde{\Gamma}^{\text{Ge}}_{\lambda\lambda'}.$

The matrix ${\bf T}$ is defined in terms of the perturbation matrix ${\bf V}$ and the perturbed Green's function ${\bf G}^+$ as
${\bf T}={\bf V} + {\bf V}{\bf G}^+ {\bf V}$,
which, after some algebraic manipulations using the orthogonality and completeness of the eigenstates, can be expressed in terms of the unperturbed 
Green's function ${\bf g}^+(\omega^2)$ as 
\bea
{\bf T}(\omega^2) = [{\bf I} - {\bf V}{\bf g}^+(\omega^2)]^{-1}{\bf V}.
\label{Tmat}
\eea
The integral form of the unperturbed Green's function ${\bf g}^+(\omega^2)$ is given by
\bea
{\bf g}^+_{ij}(\omega^2) = \lim_{z\to \omega^2+i0} \sum_\lambda \f{\langle {i}|{\lambda} \rangle \langle {\lambda}|{j}\rangle}{z-\omega^2_{\lambda}},
\label{Eq:green}
\eea
where $|{\lambda} \rangle$ are the eigenstates of the infinite unperturbed lattice and $|{i}\rangle$ is a local displacement of the $i$th degree of freedom in the direct lattice. 
For the numerical computation of ${\bf g}^+(\omega^2)$ in Eq.~\ref{Eq:green}, we have employed the tetrahedron approach of Lambin and Vigneron\cite{Lambin}. The total scattering rate due to the nanoparticles, appearing as the third term on the right hand side of Eq.~\ref{Eq:tau0}, is efficiently computed using the optical theorem \cite{Mingo10}:
\bea
1/\tau_{\lambda}^{np}=\sum_{\lambda'}{\Gamma_{\lambda\lambda'}}={\Omega\over2\omega^2}{f_{\text{np}}\over V_{\text{np}}}\text{Im}\big[\langle \lambda|\bf{T}(\omega^2)|\lambda\rangle\big]
\label{Eq:optical}
\eea

Most often in the literature, where the Born approximation is employed, the T
matrix is replaced by the perturbation matrix V.
This is justified by the expansion ${\bf T}\simeq \bf V + VgV + ...$, valid for small perturbations. In this approximation 
one obtains
\bea
\frac{1}{\tau_{\lambda}^{np}} &=& \frac{\Omega}{16~\pi^2}~g_2 \omega_{\lambda}^2~D_{\lambda}^S, \nn \\
D_{\lambda}^S &=& \sum_{\lambda'}  \Big | \sum_{k'} {\bf e}_{k'}^{\lambda}.{\bf e}_{k'}^{\lambda'^*} \Big |^2  | S_{\Delta {\bf q}} |^2~\delta(\omega_{\lambda'} - \omega_{\lambda})
\label{Born} 
\eea
 Here, $g_2=f_p \big(1-M'/M \big)^2$, $D_{\lambda}^S$ is like a phonon density of states but weighted by the structure
 factor for the nanoparticle: $S_{\Delta {\bf q}}={1\over N_{\text{p}}^2}\sum_{l\in\text{np}}e^{i{\bf R}_l\cdot {\Delta {\bf q}}}$, $N_p$ being the number of unit cells making
 up the nanoparticle. The $l'$ sum is only over those unit cells of the virtual crystal
 occupied by the atoms of the nanoparticle and $\Delta {\bf q} = {\bf q} - {\bf q'}$. The ${\bf e}_k^{\lambda}$ are phonon
 eigenvectors for the $k$ th atom in a unit cell. In the limit of a single atom impurity, Eq.~\ref{Born}
 correctly reduces to the form derived previously by Tamura for the scattering rate of
 isotope impurities in cubic crystals \cite{Tamura}.

A consequence of the Born approximation is that the sign of the perturbation does not matter: a given percent of either increase or decrease of the scatterer's mass density with respect to the host's should produce the same result. This is not true when the exact T matrix is employed, and large differences can occur with respect to the Born approximation for large mass difference, as our results show below.

\section{Results and discussion}
We have first computed the thermal conductivity of $\text{Si}_{0.5}\text{Ge}_{0.5}$ at 300K and 800K.  
The perturbative approach for isotopic impurities \cite{Ward09}, where $g_2 \sim 10^{-4}$, has for decades also been used for alloys \cite{Abeles}, where one might question its validity since $g_2$ in the alloy is several orders of magnitude larger (for $\text{Si}_{0.5}\text{Ge}_{0.5},~g_2=0.2$).  We have compared the $\text{Si}_{0.5}\text{Ge}_{0.5}$ scattering rates from Eq.~(\ref{Born}) with those obtained using the full T-matrix method, and we find these to be close, as are the alloy thermal conductivities: $\kappa_{300K}$ = 10.62 (T-matrix) vs. 10.27 (Born) W/m-K and  $\kappa_{800K}$ = 6.07 (T-matrix) vs. 6.0 (Born) W/m-K. Note that the alloy thermal conductivities are far lower than those of either bulk Si or bulk Ge because the alloy scattering is much stronger than the three-phonon scattering.  As a result, $\Delta_{\lambda}$ in Eqs.~(\ref{taulambda}) and (\ref{deltalambda}) is small and $\tau_{\lambda} \approx \tau^0_{\lambda}$.
These values are about 30\% larger than the experimental values at the same concentration and temperatures ($\sim 7.5-8$W/m-K at room temperature, $\sim 4.5-5$ W/m-K at 800K)\cite{Abeles,SlackHussain,Yonenaga}. 
There are several reasons for this. First, as already discussed at length in section II, our neglect of differences in force constants may lead to somewhat higher values. Also, experimental samples contain a certain amount of impurities and defects which also lower the thermal conductivity. In addition, there is a considerable spread in experimental results from different sources, which further attests to the various unknown factors present in experimental measurements of alloy samples. 
Finally, a recent first principles calculation of SiGe alloy thermal conductivity showed that the virtual crystal approach slightly underestimates phonon scattering in alloys \cite{Marzari}.

\begin{figure}[t]
\includegraphics[width=8.5 cm]{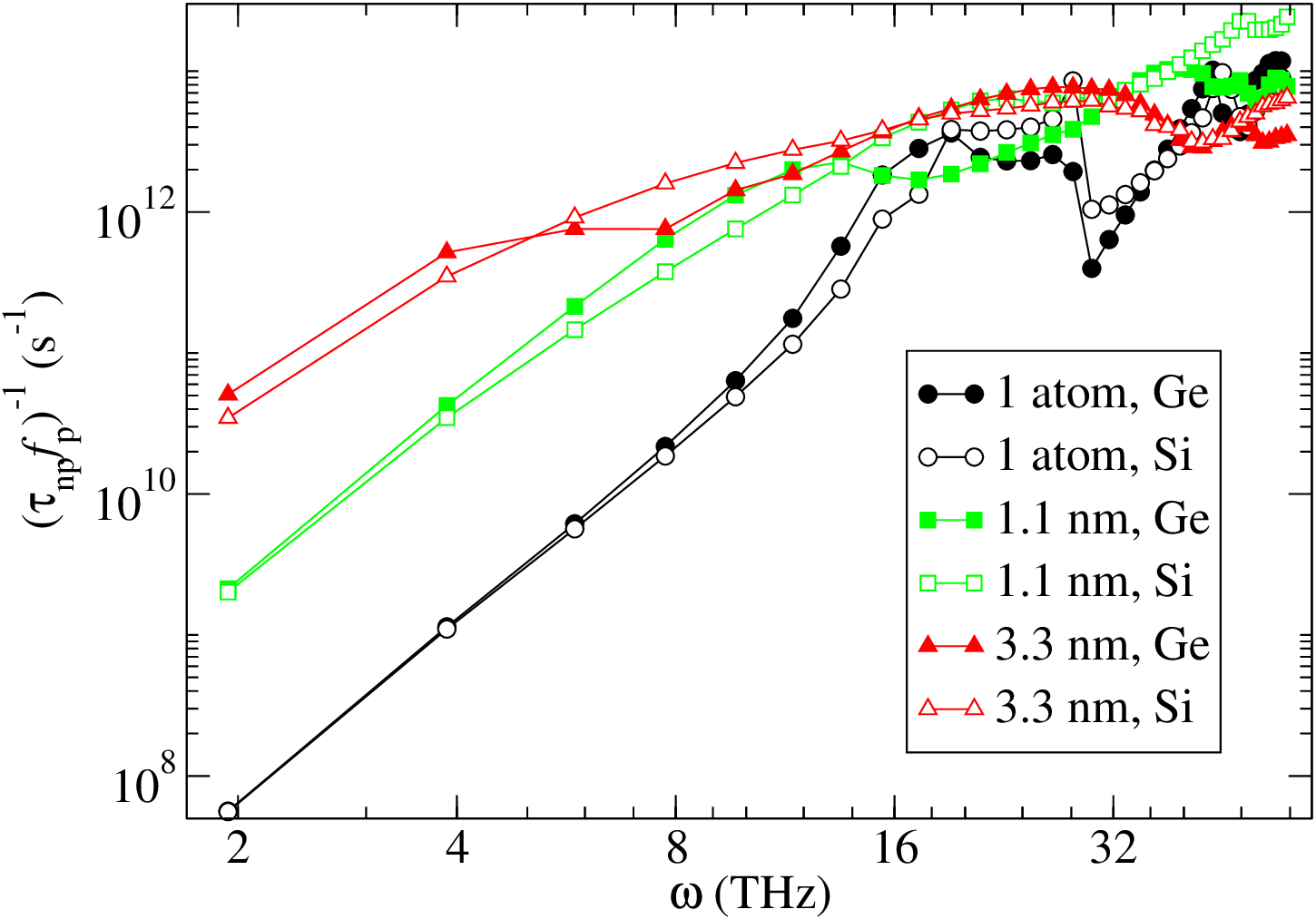}
\caption{Comparison of the scattering rates $1/\tau_{\lambda}^{np}$ normalized by nanoparticle volume fraction $f_{p}$, due to Ge and Si nano particles of different size, in a Si$_{0.5}$Ge$_{0.5}$ matrix. The curves shown correspond to the LA phonon branch along direction (100).}
\label{kappaA}
\end{figure}

Our calculation shows that for single atom scatterers, low frequency phonons are well described by the Born approximation, with almost no difference between Si and Ge impurities. This is clearly seen in figure 1, for the case of the longitudinal acoustic branch. Although, the scattering rate of Ge and Si single atoms differ importantly at high frequency, however this does not lead to much difference in $\kappa$, since $\kappa$ is dominated by low frequency phonons. It is only when we consider larger nanoparticles that differences become appreciable in the thermal conductivity. For nanoparticles of diameter 1.1 nm, containing 38 atoms, scattering rates differ considerably between the Si and Ge cases, already for $\omega$ above 5 THz. This leads to a difference between the thermal conductivities of the corresponding composites. The difference becomes even more appreciable for larger nanoparticles (see Fig.~1). Fig.~\ref{kappaA} compares the Born approximation result, with the exact scattering rates $1/\tau^{np}_{\lambda}$ for both a Si nano particle in $\text{Si}_{0.5}\text{Ge}_{0.5}$ and a Ge nano particle in $\text{Si}_{0.5}\text{Ge}_{0.5}$ for  a diameter of 3.3 nm, when the incident phonon direction is (100). 

\begin{figure}[t]
\includegraphics[width=8.5 cm]{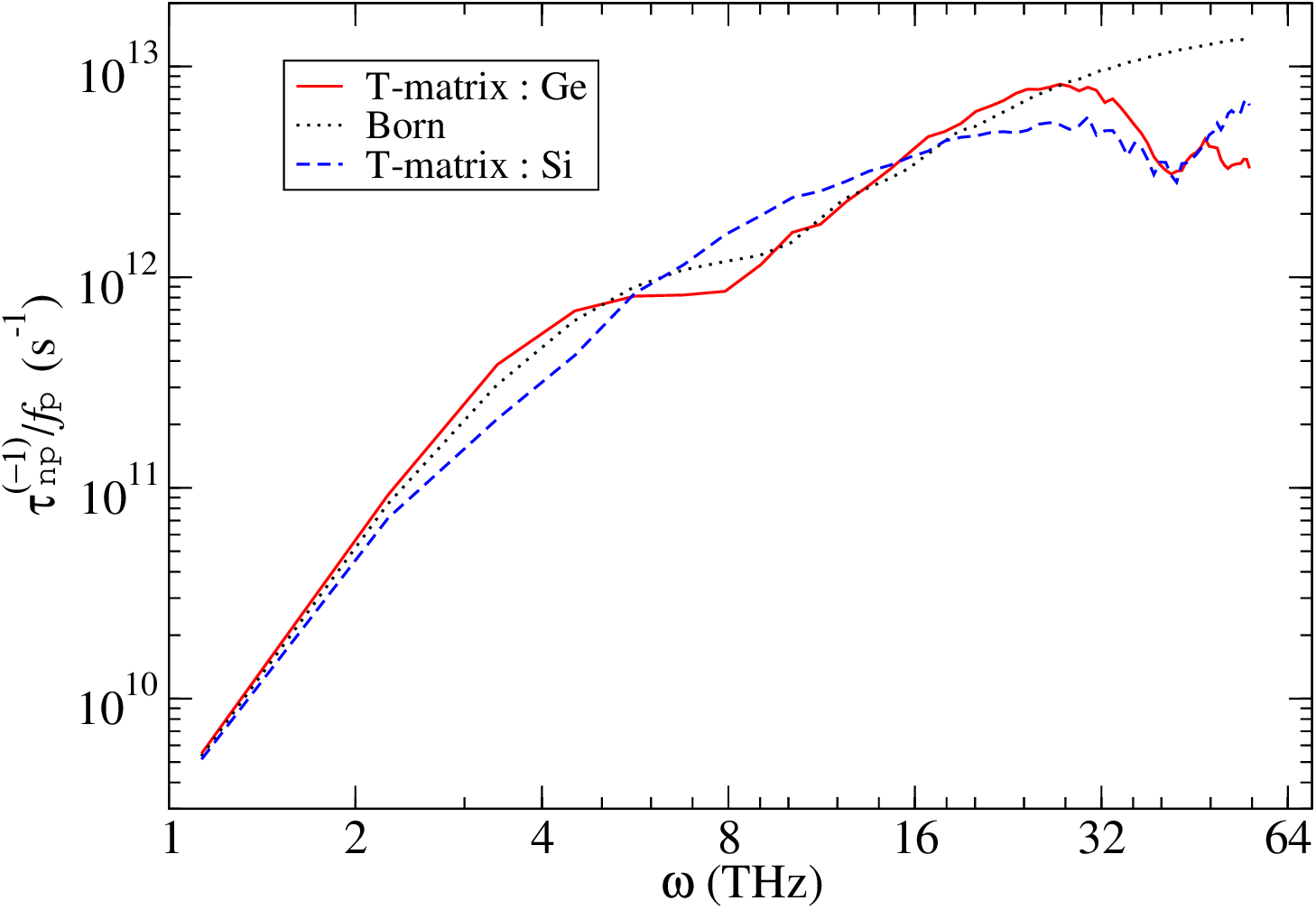}
\caption{Scattering rate $1/\tau_{\lambda}^{np}$ due to Ge and Si nano particle of diameter 3.3 nm, in a Si$_{0.5}$Ge$_{0.5}$ matrix, normalized by nanoparticle volume fraction $f_{p}$. The black dotted line is the Born approximation result from Eq.(\ref{Born}). The curves correspond to the LA phonon branch along direction (100).}
\label{kappaA}
\end{figure}

\begin{figure}[t]
\includegraphics[width=7.5 cm]{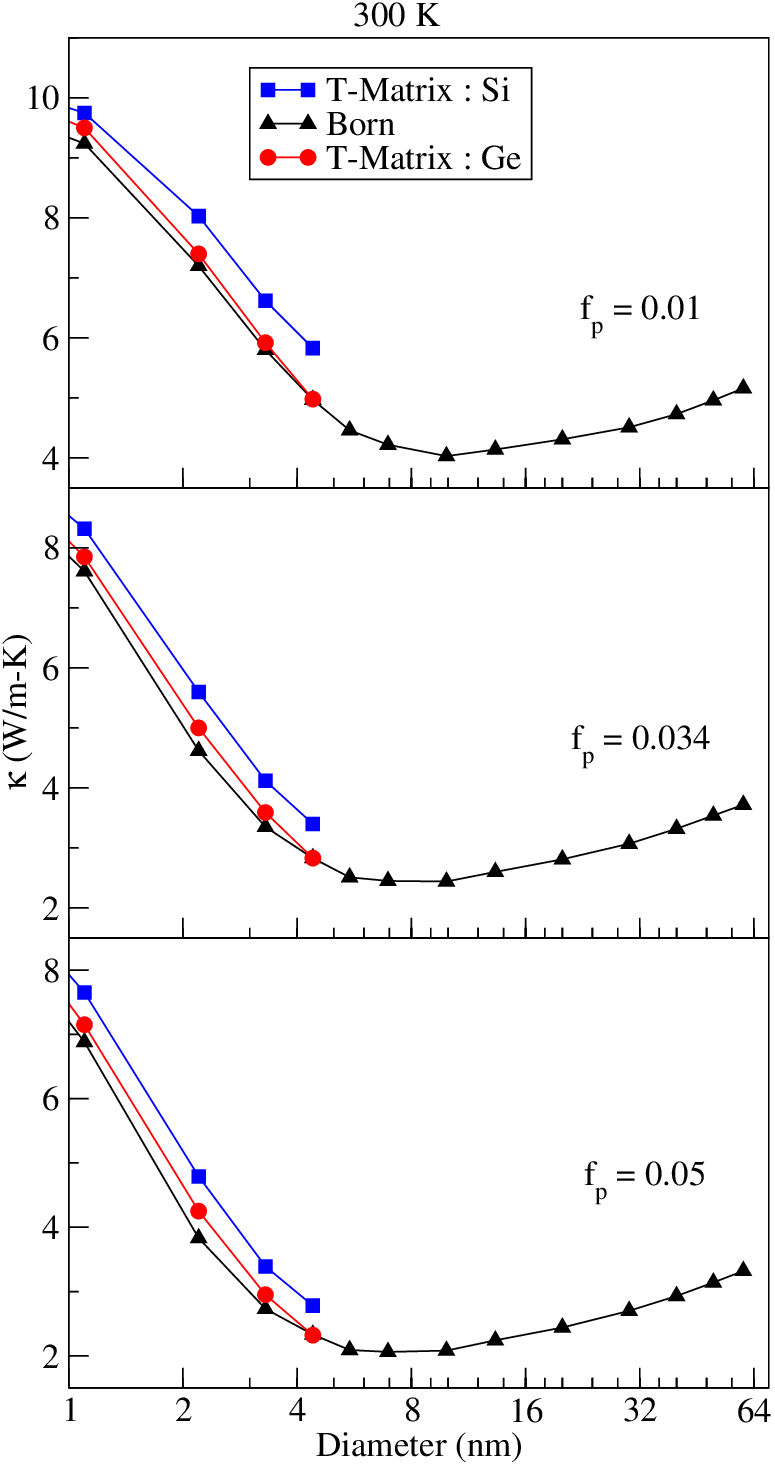}
\caption{Thermal conductivity vs. nanoparticle diameter at 1\% (top), 3.4\% (middle), and 5\% (bottom) nanoparticle concentrations, for temperature 300K. Triangles: Born+geometrical interpolation. Squares: T matrix calculation for Si nanoparticle. Circles: T matrix calculation for Ge nanoparticle.}
\label{kappaB}
\end{figure}

\begin{figure}[t]
\includegraphics[width=8.5 cm]{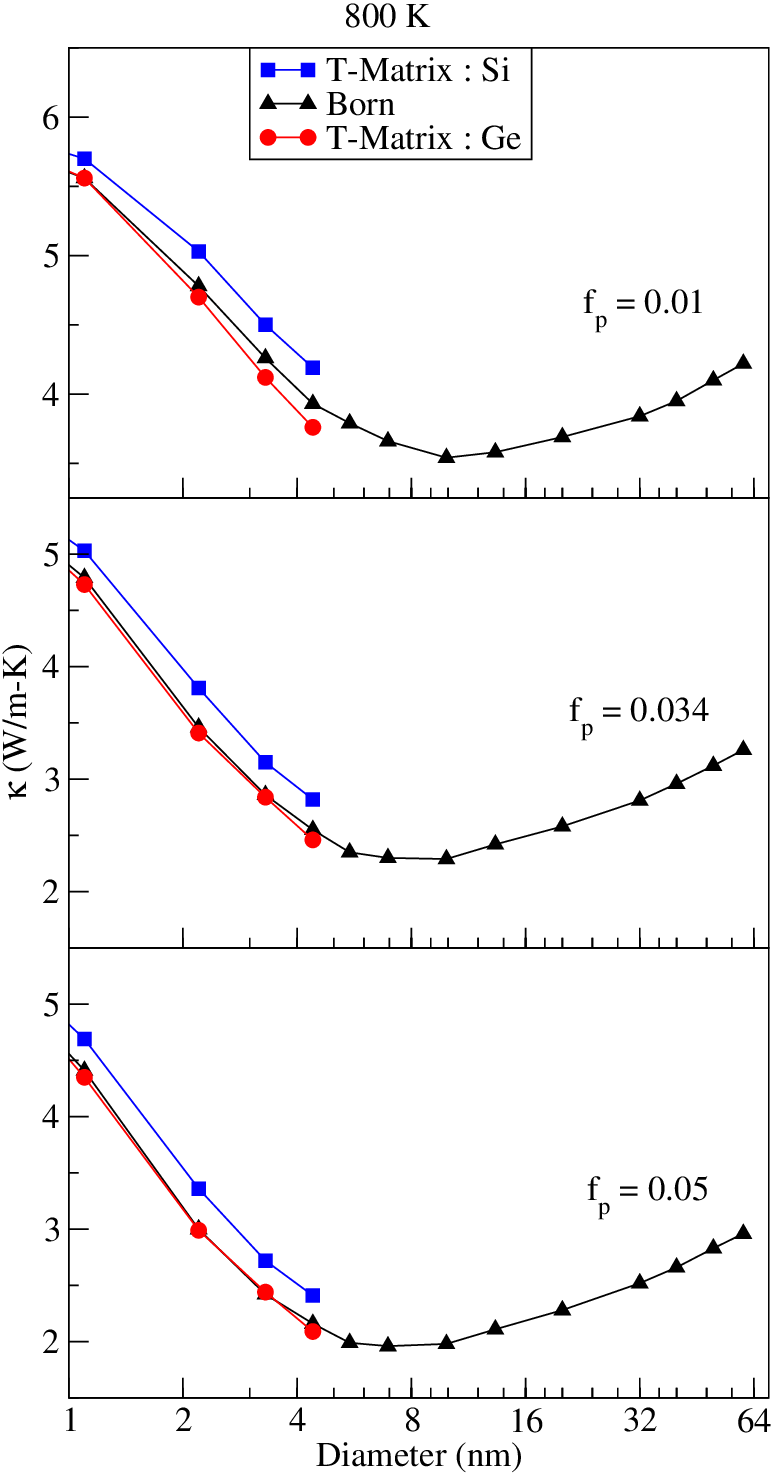}
\caption{Thermal conductivity vs. nanoparticle diameter at 1\% (top), 3.4\% (middle), and 5\% (bottom) nanoparticle concentrations, for temperature 800K. Triangles: Born+geometrical interpolation. Squares: T matrix calculation for Si nanoparticle. Circles: T matrix calculation for Ge nanoparticle.}
\label{kappaC}
\end{figure}

At high frequency both the Si and Ge cases deviate considerably from the Born approximation result (also shown in Fig.~\ref{kappaA} ). 
This occurs when the wavelengths become comparable to the size of the scatterers, so we are no longer in the Rayleigh regime. An earlier interpolation formula had been proposed to link between the Born and geometric regime scattering cross sections\cite{KimMajumdar,Schwartz,Vandersande,Joshi,Majumdar}: $1/\sigma\simeq 1/\sigma_{\text{geom}}+1/\sigma_{\text{Born}}$. Fig.~\ref{kappaB} and \ref{kappaC} show $\kappa$ versus nanoparticle size obtained using this approximated interpolation formula. There is an optimal nanoparticle size that minimizes thermal conductivity at a given concentration. These ab-initio curves confirm the simpler model predictions in Ref.~\onlinecite{MingoNanolett}, yielding a minimum for a diameter of a few nm, and a slow increase after that.

The interpolated expression still makes use of the Born approximation, so it does not inform us of possible differences between heavier and lighter scatterers. We have compared those results with the ones obtained using the T-matrix computed scattering rates. 
The plot shows quantitative differences, but the trends are the same. The T-matrix approach is very computationally demanding: a 4.4 nm diameter particle, containing 2122 atoms is already at the limit of our computing capability. Therefore, we cannot assess the exact position of the minimum for the Ge or Si nanoparticle cases, although the graphs suggest that it may take place at a diameter between 5-10 nm.

\begin{figure}[t]
\includegraphics[width=7.5 cm]{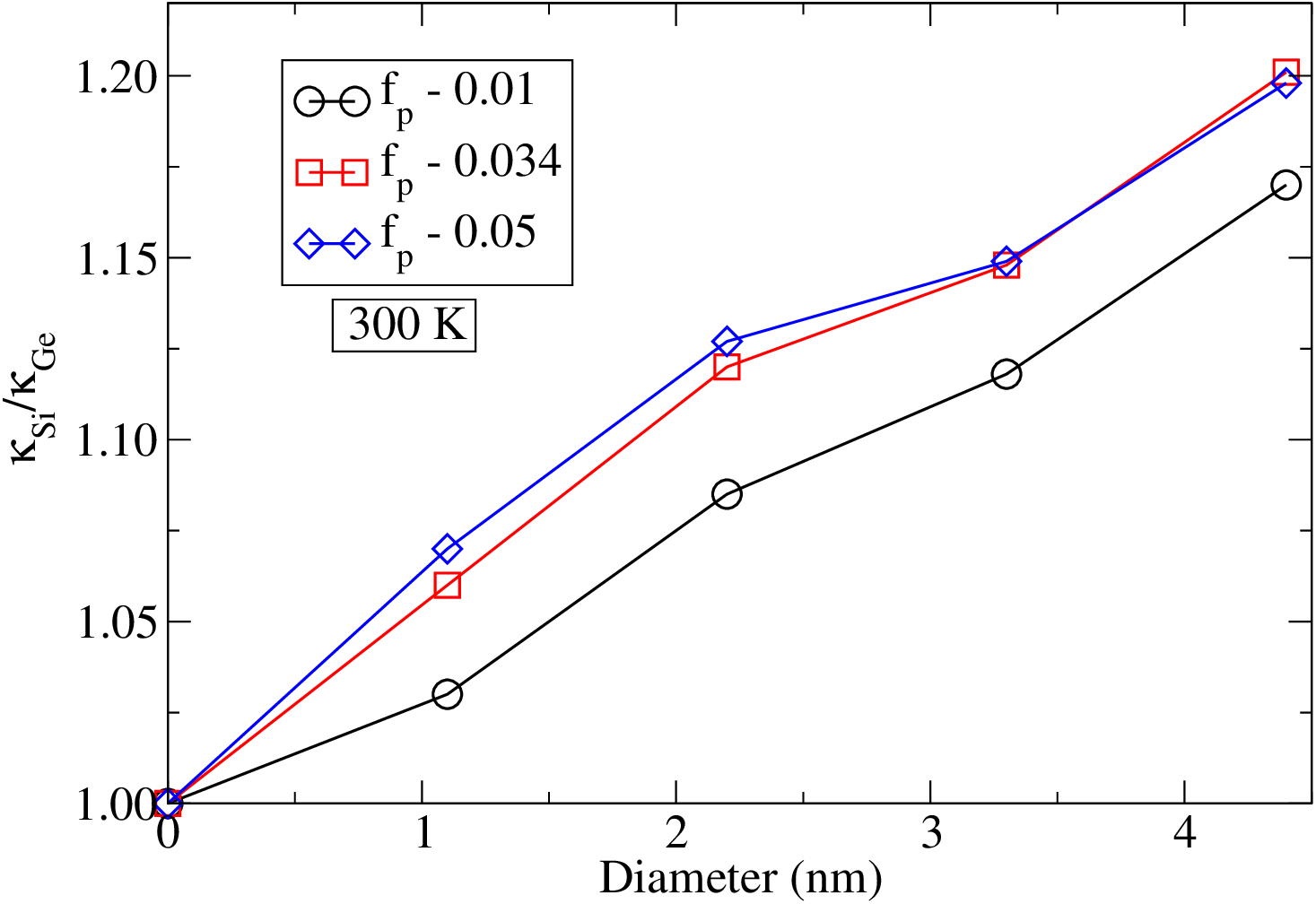}
\caption{The ratio $\kappa_{Si}/\kappa_{Ge}$, of the thermal conductivity of a SiGe matrix with embedded Si nanoparticles, $\kappa_{Si}$, to the thermal conductivity of a SiGe matrix with embedded Ge nanoparticles, $\kappa_{Ge}$, is shown as a function of nanoparticle diameter, at 300K, for three different nanoparticle concentrations, $f_{p}$. 
}
\label{kappaD}
\end{figure}

At equal nanoparticle size and concentration, the calculated $\kappa$ is always smaller for Ge (heavier) than for Si (lighter) nanoparticles. Their density difference with respect to Si$_{0.5}$Ge$_{0.5}$ is the same except for the sign, so in the framework of the Born approximation they should show an identical effect. This is clearly not the case, as shown in Fig.~\ref{kappaD}. This figure shows the ratio between the conductivities of the two cases, as a function of nanoparticle diameter: the thermal conductivity of the Ge nanoparticle case can be up to 20\% lower than that of the Si nanoparticle case. 
The difference between the two cases highlights the very different densities of the particles and the medium, close to 20\%. The fact that Ge nanoparticles affect $\kappa$ more than Si nanoparticles is directly linked to their higher scattering rate at low frequency, visible in Fig.~\ref{kappaA}. The low frequencies are the ones that make the largest contribution to $\kappa$. This is because high frequencies already have very short mean free paths, and so most of the heat in the alloy is carried by low frequency phonons. Thus, even though Si displays higher scattering rates at some intermediate and higher frequencies, it is the small scattering rates at low frequencies that determine the difference between the lighter and heavier types of nanocomposite.

A qualitative difference in the scattering rates of lighter and heavier impurities had been shown as early as 1963 for a model FCC scalar lattice with single and double impurities\cite{Takeno}. Our results for the SiGe problem display the same kind of behavior, where the heavy impurities scatter more strongly than the light ones at low frequency. This can be qualitatively understood by making an analogy with electron scattering by a local potential. Heavier impurities are analogous to a potential well, whereas lighter ones are analogous to a potential barrier. From elementary scattering theory, the low frequency scattering cross section of a potential well is larger than that of a potential hump, in agreement with the trend observed. Nevertheless, to our knowledge, the effect of arbitrarily sized nanoparticles on a realistic 3 dimensional system has not previously been quantitatively investigated, and its consequences on thermal conductivity have not been addressed. 

Some further comments are in order. In principle, the techniques presented here would also allow us to study other nanoparticle shapes and compositions. Our choice of pure Si and Ge spherical nanoparticles has been motivated by simplicity. Experimentally, it may prove difficult to embed such nanoparticles into a SiGe matrix, because Si and Ge are fully miscible. Although high concentration Ge nanoparticles with flat pyramidal or hemispherical shapes have been grown inside a Si \cite{Pernot} and SiGe matrix \cite{DavidH} in the past, for the sake of clarity we have avoided introducing any experimentally determined morphological characteristics in our calculation.

\section{Conclusions}
We have preformed a parameter free first principles calculation of the thermal conductivity of SiGe alloys with embedded Si or Ge nanoparticles. In contrast with the commonly used Born approximation, it is found that embedding nanoparticles in the material affects its thermal conductivity differently depending on whether the nanoparticles are relatively heavier or lighter than the embedding matrix. The calculation predicts that heavier nanoparticles (Ge) should be more efficient than lighter ones (Si) in reducing the $\kappa$ of Si$_{0.5}$Ge$_{0.5}$. This behavior is determined by the higher scattering rate for heavier nanoparticles at low frequency, which is not predicted by the standard Born approximation, but is captured by the full Green's function calculation. Nevertheless, the approximated Born + geometrical approximation is found to work remarkably well, being within 20\% of the exact result. The ab initio calculation also confirms the existence of an optimal nanoparticle size that minimizes thermal conductivity, which had been previously predicted using a simpler model\cite{MingoNanolett}.

\section*{Acknowledgements}
We acknowledge support from Fondation Nanosciences, Agence Nationale de la Recherche, and the National Science Foundation. NM thanks Ali Shakouri for helpful discussions.


\begin{thebibliography}{}
\bibitem{KimShakouri}
W. Kim, J. Zide, A. Gossard, D. Klenov, S. Stemmer, A. Shakouri, and A. Majumdar, Phys. Rev. Lett. {\bf 96}, 045901, (2006).

\bibitem{Zebarjadi}M. Zebarjadi, K. Esfarjani, Z. Bian, and A. Shakouri, Nano Lett. {\bf 11}, 225 (2011).

\bibitem{Kanatzidis}
K. F. Hsu, S. Loo, F. Guo, W. Chen, J. S. Dyck, C. Uher,T. Hogan, E. K. Polychroniadis and M. G. Kanatzidis, Science {\bf 303}, 818, (2004).



\bibitem{Klemens}
L. A. Turk and P. G. Klemens, Phys. Rev. B {\bf 9}, 4422, (1974).

\bibitem{SlackHussain}
 G. A. Slack and M. A. Hussain, J. Appl. Phys. 70, 2694, (1991).

\bibitem{KimMajumdar} 
W. Kim, and A. Majumdar, J. Appl. Phys. {\bf 99}, 084306, (2006).

\bibitem{MingoNanolett}
 N. Mingo, D. Hauser, N. P. Kobayashi, M. Plissonnier, and A. Shakouri, Nano Letters {\bf 9}, 711, (2009).

\bibitem{Sparavigna}
M. Omini and A. Sparavigna, Nuovo Cimento D {\bf 19}, 1537 (1997).

\bibitem{Broido07}
D. A. Broido, M. Malorny, G. Birner, N. Mingo and D. A. Stewart, Appl. Phys. Lett. {\bf 91}, 231922, (2007).

\bibitem{Ward09}
A. Ward, D. A. Broido, D. A. Stewart, G. Deinzer, Phys. Rev. B {\bf 80}, 125203, (2009).

\bibitem{Lindsay}
L. Lindsay, D. A. Broido, and N. Mingo,Phys. Rev. B {\bf 82}, 161402(R), (2010).

\bibitem{Ziman}
J.M. Ziman, Electrons and Phonons (Clarendon Press, London, 1962).

\bibitem{Abeles}
B. Abeles, Phy. Rev. {\bf 131}, 1906, (1963).

\bibitem{Klemens55}
P. G. Klemens, Proc. Phys. Soc. A {\bf 68}, 1113 (1955).
\bibitem{Mingo10}
N. Mingo, K. Esfarjani, D. A. Broido, and D. A. Stewart, Phys. Rev. B {\bf 81}, 045408, (2010).

 
\bibitem{Lambin}  
Ph. Lambin and J. P. Vigneron, Phys. Rev. B {\bf 29}, 3430 (1984). 

\bibitem{Tamura}
S. I. Tamura, Phys. Rev. B , {\bf 27}, 858, (1983).

\bibitem{Yonenaga} I. Yonenaga, T. Akashi, and T. Goto, J. Phys. Chem. Sol. {\bf 62} (2001) 1313.

\bibitem{Marzari} J. Garg, N. Bonini, B. Kozinsky, and N. Marzari, Phys. Rev. Lett., {\bf 106}, 045901 (2011).


\bibitem{Schwartz} J. W. Schwartz and C. T. Walker, Phys. Rev. {\bf 155}, 969 (1967) 

\bibitem{Vandersande} J. W. Vandersande, Phys. Rev. B {\bf 15}, 2355 (1977) 

\bibitem{Joshi} Y. P. Joshi, Phys. Stat. Sol. (b) {\bf 95}, 627 (1979).

\bibitem{Majumdar}
A. Majumdar, ASME Trans. J. Heat Transfer {\bf 115}, 7  (1993).



\bibitem{Takeno}
S. Takeno, Prog. of Theo. Phys. , {\bf 29}, No. 2, 191, (1963).

\bibitem{Pernot}
G. Pernot {\it et al.}, Nature Materials, {\bf 9}, 491 (2010).

\bibitem{DavidH}
D. Hauser, M. Plissonnier, L. Montes, J. Simon, and G. Savelli, 
to be published.

\end{thebibliography}
\end{document}